\documentclass[preprint,review,12pt]{elsarticle}

\usepackage{lineno,hyperref}
\modulolinenumbers[5]

\journal{Acta Materialia}

\usepackage{graphicx}
\usepackage{color}

\usepackage{amssymb}
\usepackage{amsmath}

\begin{document}

\begin{frontmatter}

\title{Study of the thermochromic phase transition in CuMo$_{1-x}$W$_x$O$_4$ solid solutions at the W L$_3$-edge by resonant X-ray emission spectroscopy  }

\author[ISSP]{Inga Pudza\corref{ip}}
\cortext[ip]{Corresponding author}
\ead{inga.pudza@cfi.lu.lv}

\author[ISSP,UP]{Aleksandr Kalinko}
\ead{aleksandr.kalinko@desy.de}

\author[ISSP]{Arturs Cintins}
\ead{arturs.cintins@cfi.lu.lv}

\author[ISSP]{Alexei Kuzmin\corref{ak}}
\cortext[ak]{Corresponding author}
\ead{a.kuzmin@cfi.lu.lv}

\address[ISSP]{Institute of Solid State Physics, University of Latvia, Kengaraga street 8, LV-1063 Riga, Latvia}

\address[UP]{Department of Chemistry and Center for Sustainable Systems Design, Paderborn University, 33098 Paderborn, Germany}

\begin{abstract}
Polycrystalline CuMo$_{1-x}$W$_x$O$_4$ solid solutions were studied by resonant X-ray emission spectroscopy (RXES) at the W L$_3$-edge to follow a variation of the tungsten local atomic and electronic structures across thermochromic phase transition as a function of sample composition and temperature. The experimental results were interpreted using ab initio calculations.  The crystal-field splitting parameter $\Delta$ for the 5d(W)-states was obtained from the analysis of the RXES plane and was used to evaluate the coordination of tungsten atoms. 
Temperature-dependent RXES measurements were successfully employed to determine the hysteretic behaviour of  the structural phase transition between the $\alpha$ and $\gamma$ phases in CuMo$_{1-x}$W$_x$O$_4$ solid solutions on cooling and heating, even at low ($x < 0.10$) tungsten content.  
It was found that tungsten ions have octahedral coordination for $x > 0.15$ in the whole studied temperature range (90-420~K), whereas their coordination changes from tetrahedral to octahedral upon cooling for smaller ($x \leq 0.15$) tungsten content. Nevertheless, some amount of tungsten ions
was found to co-exists in the octahedral environment at room temperature for $x < 0.15$.  The obtained results correlate well with the color change in these solid solutions. 
\end{abstract}


\end{frontmatter}


\newpage

\section{Introduction}
\label{intro}

CuMo$_{1-x}$W$_x$O$_4$ solid solutions represent a class of functional materials demonstrating a wide range of physical and chemical properties such as thermochromic \cite{Gaudon2007a,Gaudon2007b,Gaudon2009,Yanase2013,Robertson2015,Wu2020}, piezochromic \cite{Gaudon2007b,Robertson2015,Blanco2015}, halochromic \cite{Gaudon2010}, photoelectrochemical \cite{Hill2013}, thermosalient \cite{Robertson2015} and catalytic \cite{Liang2018}. 

\begin{figure}[t]
	\centering
	\includegraphics[width=0.9\textwidth]{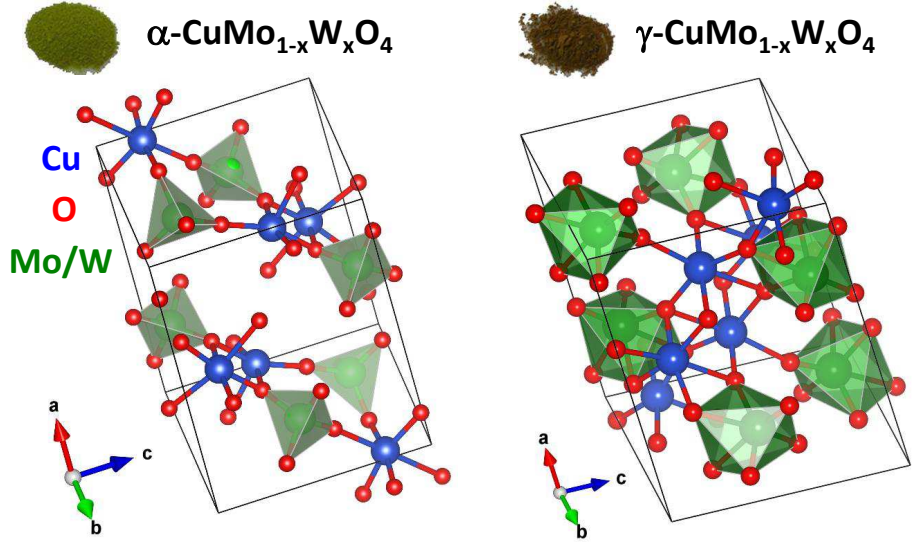}
	\caption{Crystal structures and unit cells of high-temperature green ($\alpha$) and low-temperature brownish-red ($\gamma$) CuMo$_{1-x}$W$_x$O$_4$ phases \protect\cite{Robertson2015,Wiesmann1997}. MoO$_4$ and MoO$_6$ polyhedra are indicated in the $\alpha$- and $\gamma$-phases, respectively. Small red balls are oxygen atoms, medium-sized blue balls are copper atoms, and large green balls are molybdenum/tungsten atoms. Photographs of the two phases are also shown.}
	\label{fig1}
\end{figure}

The strongest thermochromic effect is observed on cooling for small tungsten content ($x < 0.15$), when the change of the material color from green to brown is due to the first-order phase transition caused by the Mo(W) displacement from the tetrahedral site (as in $\alpha$-CuMoO$_4$) to the octahedral site (as in  $\gamma$-CuMoO$_4$) (Fig. \ref{fig1}) \cite{Gaudon2007a, Robertson2015,Joseph2020}. The similar color change occurs also in $\alpha$-CuMoO$_4$ on heating  \cite{Steiner2001}, however, in this case, no phase transition takes place, the coordination of metal ions remains unchanged, and the variation of color is due to the lattice expansion and strong enhancement of specific thermal disorder \cite{Jonane2019}. For tungsten content above $x \sim 0.15$,  CuMo$_{1-x}$W$_x$O$_4$ solid solutions crystallize in the wolframite-type structure with the octahedral coordination of metal ions and do not manifest pronounced thermochromic behaviour. Therefore, the knowledge  of the local environment of metal ions in these mixed compounds is crucial for understanding and controlling their properties. 
   
Conventionally, X-ray absorption spectroscopy (XAS) is used to probe  the local environment of atoms in complex materials \cite{Bordiga2013,YOUNG2014,Mastelaro2018}. However, in the case of tungstates and molybdates, often composed from strongly distorted polyhedra, the recovery of structural information from XAS requires the use of a computationally heavy and time-consuming approach based on reverse Monte Carlo (RMC) simulations \cite{Timoshenko2015, Kalinko2016, Timoshenko2016, Kuzmin2020}. Moreover, we have previously shown that a reliable structural analysis requires the knowledge of X-ray absorption spectra at least for all metal edges \cite{Timoshenko2015, Jonane2020}. In the case of solid solutions,  the low content of one of the component limits the quality of the experimental data, which additionally complicates the analysis. Finally, the close values of the metal--oxygen interatomic distances in the radial distribution functions (RDFs) obtained using the RMC simulation make it often difficult to unambiguously identify the atoms belonging to the first coordination shell of metal and, as a result, the type of its coordination polyhedron. 

In tungsten oxides and related materials, the splitting of the tungsten 5d band depends on the crystal field of ligands and can be experimentally probed by the W L$_3$-edge XAS \cite{Yamazoe2008,Jayarathne2014}, that can be used to distinguish between different local symmetries of the tungsten coordination polyhedron.
In particular, the tungsten 5d-band splits into 5d(t$_{2g}$) and 5d(e$_g$) sub-bands in octahedral and 5d(t$_2$) and 5d(e) sub-bands in tetrahedral coordination.
However, the small value of the crystal field splitting (about several electron volts) and the large value of the natural width of the excited 2p$_{3/2}$(W) level ($\sim$4.5~eV \cite{KESKIRAHKONEN1974}) often mask the difference between the two coordinations \cite{BALERNA1991,Kuzmin1997spie,Charton2002}. Therefore, a more sophisticated experimental approach should be used to more reliably address this issue. 

In this study, we used the resonant X-ray emission spectroscopy (RXES) at the W L$_3$-edge to determine a variation of the tungsten local atomic and electronic structure in CuMo$_{1-x}$W$_x$O$_4$  solid solutions across thermochromic phase transition as a function of sample composition and temperature. We demonstrate that the analysis of the
W L$_3$-edge white line splitting provides a robust tool for distinguishing tetrahedral and octahedral coordinations of tungsten ions in these solid solutions.

\section{Experimental details}
\label{exper}
Polycrystalline CuMo$_{1-x}$W$_x$O$_4$ powders were synthesized using a solid-state reaction method by heating a mixture of CuO and MoO$_3$ powders with a stoichiometric amount of WO$_3$ at 650$^\circ$C in air for 8 hours followed by cooling down to the room temperature. One group of samples with $x\leq 0.15$ was greenish, and the second group with $x \geq 0.20$ was brownish.
All samples were characterized by X-ray powder diffraction and micro-Raman spectroscopy (see in the Supplementary Information). 

The temperature-dependent RXES experiments were performed at the P64 Advanced X-Ray Absorption Spectroscopy beamline of the PETRA III (HASYLAB/DESY) storage ring using recently constructed RXES endstation \cite{PETRAP64,HAMOS2020}. The X-ray beam from an undulator was monochromatized by a liquid-nitrogen-cooled double-crystal Si(311) monochromator and focused in vertical and horizontal planes (100~$\mu$m $\times$ 240~$\mu$m) by two mirrors with Rh coating positioned downstream of the monochromator. 
The high stability and high intensity (5$\cdot$10$^{11}$ photons/s) of the P64 beamline provided the  possibility to accumulate high quality/resolution RXES data. The ionization chamber was used for incident X-ray intensity monitoring. The XES signal was dispersed by von Hamos-type spectrometer equipped with cylindrical bent Si(444) analyser crystals  \cite{HAMOS2020} and collected by Dectris 2D Pilatus 300K detector. 
The monochromator was calibrated at the W L$_3$-edge using tungsten foil and
the energy scale of the spectrometer was set according to the position of  the elastic line on the detector. 
The inelastically scattered X-ray signal and  W L$\alpha_1$ and L$\alpha_2$  fluorescence were acquired in the RXES plane. The high-resolution W L$_3$-edge XANES (HERFD-XANES) was extracted from the RXES plane at fixed emission energy 8398.5$\pm$0.2 eV. The temperature of the sample was controlled using liquid nitrogen (LN$_2$) cryostat Linkam THMS600 in the range of 90-450 K. All samples were measured in the form of sintered powders. 

The W L$_3$-edge XANES spectra of the same powder samples recorded at room-temperature in transmission mode at the Elettra XAFS bending-magnet beamline \cite{Di2009} were used for comparison. The experimental details can be found in \cite{Jonane2019}.

\section{FDMNES calculations}
\label{calc}

The simulation of the experimental W L$_3$-edge XANES spectra was performed based on the finite difference method (FDM) by the ab initio real-space FDMNES code \cite{Joly2001, Bunuau2009}, thus avoiding the muffin-tin potential approximation. 
The XANES spectra were calculated using real energy-dependent Hedin-Lundqvist exchange-correlation potential \cite{Hedin1971} taking into account spin-orbit interactions and the self-consistent cluster potential. The calculations were performed for 4.0~\AA\ radius clusters constructed around the absorbing tungsten atom from the crystallographic structures of interest, since the effect of the crystal field is related to the interaction between tungsten and nearest oxygen atoms.

All calculated XANES spectra were convoluted with a Lorentzian broadening function with a FWHM given by $\Gamma_{hole}$=1~eV to mimic the broadening due to the core-hole lifetime and experimental resolution. The effect of the core-hole was also examined by performing calculations for the excited (with the core hole) and non-excited (without the core hole) absorbing tungsten atom. 

The structures of $\alpha$-CuMoO$_4$ and $\gamma$-CuMoO$_4$ contain three non-equivalent sites for copper (Cu1, Cu2, Cu3) and molybdenum (Mo1, Mo2, Mo3) atoms  \cite{Wiesmann1997}. 
However, all molybdenum atoms occupy either distorted tetrahedral (in $\alpha$-CuMoO$_4$) or
distorted octahedral  (in $\gamma$-CuMoO$_4$) positions. 
In CuWO$_4$, there is only one tungsten position with distorted octahedral coordination  \cite{Forsyth1991}.

\begin{figure}[t]
	\centering
	\includegraphics[width=0.9\textwidth]{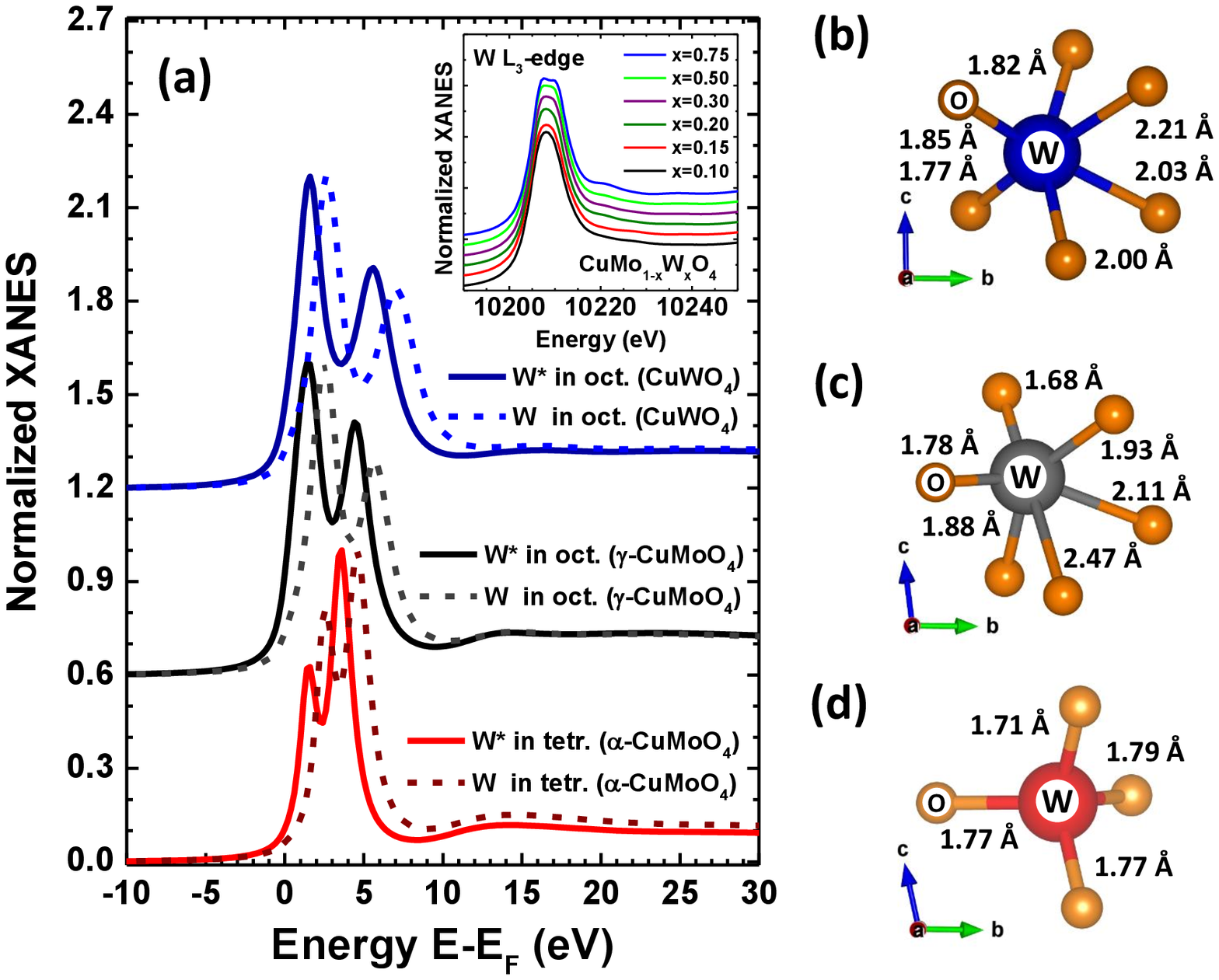}
	\caption{(a) Calculated W L$_3$-edge XANES spectra for tungsten atoms in distorted octahedral (CuWO$_4$ and W substituting molybdenum atoms at the  Mo2 site in $\gamma$-CuMoO$_4$) and tetrahedral (W substituting molybdenum atoms at the  Mo2 site  in $\alpha$-CuMoO$_4$) environments. 
    The results for the excited state with the core hole (W*, solid curves) and 
    non-excited state without the core hole (W, dashed curves) are given. 
	Tungsten--oxygen polyhedra with indicated interatomic distances in (b) CuWO$_4$,  (c) W at the Mo2 site in $\gamma$-CuMoO$_4$ and (d) W at the Mo2 site in $\alpha$-CuMoO$_4$ are also shown. 
	Inset shows the experimental W L$_3$-edge XANES spectra in CuMo$_{1-x}$W$_x$O$_4$ solid solutions measured in the transmission mode. }
	\label{fig2}
\end{figure}

The calculated W L$_3$-edge XANES spectra are shown in Fig.\ \ref{fig2} for  
CuWO$_4$ and  tungsten atom substituting molybdenum atom at the Mo2 site in  $\alpha$-CuMoO$_4$ (distorted tetrahedral environment) and $\gamma$-CuMoO$_4$ (distorted octahedral environment). The XANES spectra calculated for tungsten atoms placed at the Mo1 or Mo3 sites are close to that at the Mo2 site and are not shown. Note that all calculations were performed based on the  crystallographic structures obtained from diffraction \cite{Wiesmann1997,Forsyth1991}, without any structure relaxation due to the substitution.  

\section{Results and discussion}
\label{results}

\subsection{Crystal field splitting in the W L$_3$-edge XANES \& RXES}

At ambient conditions, CuMo$_{1-x}$W$_x$O$_4$ solid solutions exist in one of three crystal structures with triclinic $P\bar{1}$ symmetry, which are isostructural to the high-pressure phases of CuMoO$_4$  \cite{Wiesmann1997}.
In the $\alpha$-phase, $2/3$ and $1/3$ of Cu$^{2+}$ cations form distorted CuO$_6$ octahedra and CuO$_5$ square-pyramids, respectively, whereas Mo$^{6+}$ cations form MoO$_4$ tetrahedra. The crystal structure of $\gamma$-CuMoO$_4$ is built up of distorted CuO$_6$ and MoO$_6$ octahedra. Finally, CuMoO$_4$-III phase has a wolframite-type structure, in which both Cu$^{2+}$ and  Mo$^{6+}$ ions occupy octahedral sites. The sequence  ($\alpha\rightarrow\gamma\rightarrow$ III) of structural transitions upon increasing W content is similar to that under hydrostatic pressure \cite{Wiesmann1997}. Therefore, the ability to tune the functional properties of  CuMo$_{1-x}$W$_x$O$_4$  solid solutions by the introduction of tungsten ions is often explained by the effect of locally-induced chemical pressure that appears due to a small difference between the ionic radii of Mo$^{6+}$ (0.59~\AA) and W$^{6+}$ (0.60~\AA) ions \cite{Shannon1976}. Besides, the increase of the $\alpha$ $\leftrightarrow$ $\gamma$ phase transition temperature at high tungsten content was attributed to the stronger preference of tungsten ions to the octahedral environment \cite{Gaudon2007a,Gaudon2007b}.

According to the crystal-field theory \cite{Bersuker2010}, the d-orbitals of the metal ion split in the octahedral field of six oxygen atoms into two groups (triply degenerate t$_{2g}$ orbitals and doubly degenerate e$_g$ orbitals) with the energy difference $\Delta_{oct}$, where the t$_{2g}$ orbitals have lower energy than the e$_g$ ones. 
In tetrahedral coordination, the splitting of the d orbitals occurs into two groups with the energy difference $\Delta_{tet}$ between the lower energy, doubly degenerate e orbitals and the higher energy, triply degenerate t$_2$ orbitals. The splitting magnitudes for regular tetrahedral $\Delta_{tet}$ and octahedral $\Delta_{oct}$ coordinations are related as $\Delta_{tet} = 4/9 \Delta_{oct}$ for similar W--O distances \cite{Bersuker2010}. For oxidation state W$^{6+}$, all five d-orbitals are vacant, therefore, two electron transitions from 2p$_{3/2}$ to split 5d states are possible and will have different X-ray absorption intensities. For tetrahedral WO$_4$ coordination, the intensity ratio will be e:t$_2$=2:3, whereas for octahedral WO$_6$ coordination, it will be t$_{2g}$:e$_g$=3:2.      	
Thus, the knowledge of splitting value or intensity ratio for the two transitions in CuMo$_{1-x}$W$_x$O$_4$  solid solutions  allows one to conclude on the type of tungsten ion coordination. Further, we will demonstrate how the analysis of the W L$_3$-edge RXES data provides an opportunity to gain this information. 

The XANES spectra of CuMo$_{1-x}$W$_x$O$_4$  solid solutions at  the  W  L$_3$-edge  contain  a  strong resonance (see the inset in Fig.\ \ref{fig2}(a)), the  so-called  ''white line``   \cite{Brown1977},  
located  just below the continuum threshold and corresponding to  the dipole-allowed electron transition from the 2p$_{3/2}$(W) level 	to a quasi-bound 5d(W)+2p(O) mixed-state in the presence of the core hole \cite{BALERNA1991}.

Although the W L$_3$-edge XANES is sensitive to the coordination of the absorbing atom, a very short natural lifetime of the excited state with the core hole at the 2p$_{3/2}$(W) level leads to significant intrinsic broadening ($\sim$4.5~eV \cite{KESKIRAHKONEN1974}) of the absorption spectrum. As a result of the limited resolution, the crystal field splitting of the 5d(W) states is smeared out and cannot be usually resolved in the conventional transmission experiment (see the inset in Fig.\ \ref{fig2}(a)). However, some splitting of the white line maximum is  nevertheless observed in the XANES spectrum measured in the transmission mode for a high tungsten content ($x = 0.75$). 

Unlike the conventional XANES, RXES is a second-order process involving X-ray absorption and then X-ray emission described by the Kramers-Heisenberg formalism \cite{Kramers1925,Glatzel2005,Rueff2010}. Experimental resolution is governed by the lifetime of intermediate and final states
involved in the RXES process and Heisenberg's uncertainty principle. Note that the intermediate state of RXES is the same as the final state of the first-order optical process in XAS \cite{Glatzel2009,RUEFF2013,Rovezzi2014}.

To interpret the experimental data, we performed the W L$_3$-edge XANES calculations using the FDMNES code \cite{Joly2001, Bunuau2009}. As a result,  information was obtained on the splitting of the 5d(W) orbitals for three local geometries of tungsten ions similar to those in CuWO$_4$, $\alpha$-CuMoO$_4$ and $\gamma$-CuMoO$_4$ (Fig.\ \ref{fig2}).  The white line in the calculated XANES spectra is split by about 4.0~eV and 3.0~eV, respectively, in CuWO$_4$ and $\gamma$-CuMoO$_4$ (W substitutes Mo at the Mo2 site) due to the distorted octahedral crystal-field, whereas a smaller splitting of about 2.0~eV is characteristic of a tungsten ion in the tetrahedral environment, as when it is located at the Mo2 site in $\alpha$-CuMoO$_4$. 

The splitting of the 5d(W) orbitals can be observed experimentally by detecting the RXES due to 3d$_{5/2}$(W)$\rightarrow$2p$_{3/2}$(W) electron transition,  since the lifetime of the final state with a hole at the 3d$_{5/2}$(W) orbital is longer than that with a hole at the 2p$_{3/2}$(W) orbital (the W L$_{3}$-edge) \cite{Wach2020}.  
Therefore, the lifetime broadening present in conventional XANES spectra is significantly reduced in the RXES experiment. As a result, additional features become visible in the HERFD-XANES spectra, which are blurred or poorly resolved with the traditional approach.
A high-resolution spectrometer with the high brightness X-ray source and highly effective detectors are required to acquire them. At the same time, high penetration of hard X-ray radiation allows one to probe the sample bulk. 

\begin{figure*}[ht]
	\centering
	\includegraphics[width=0.95\textwidth]{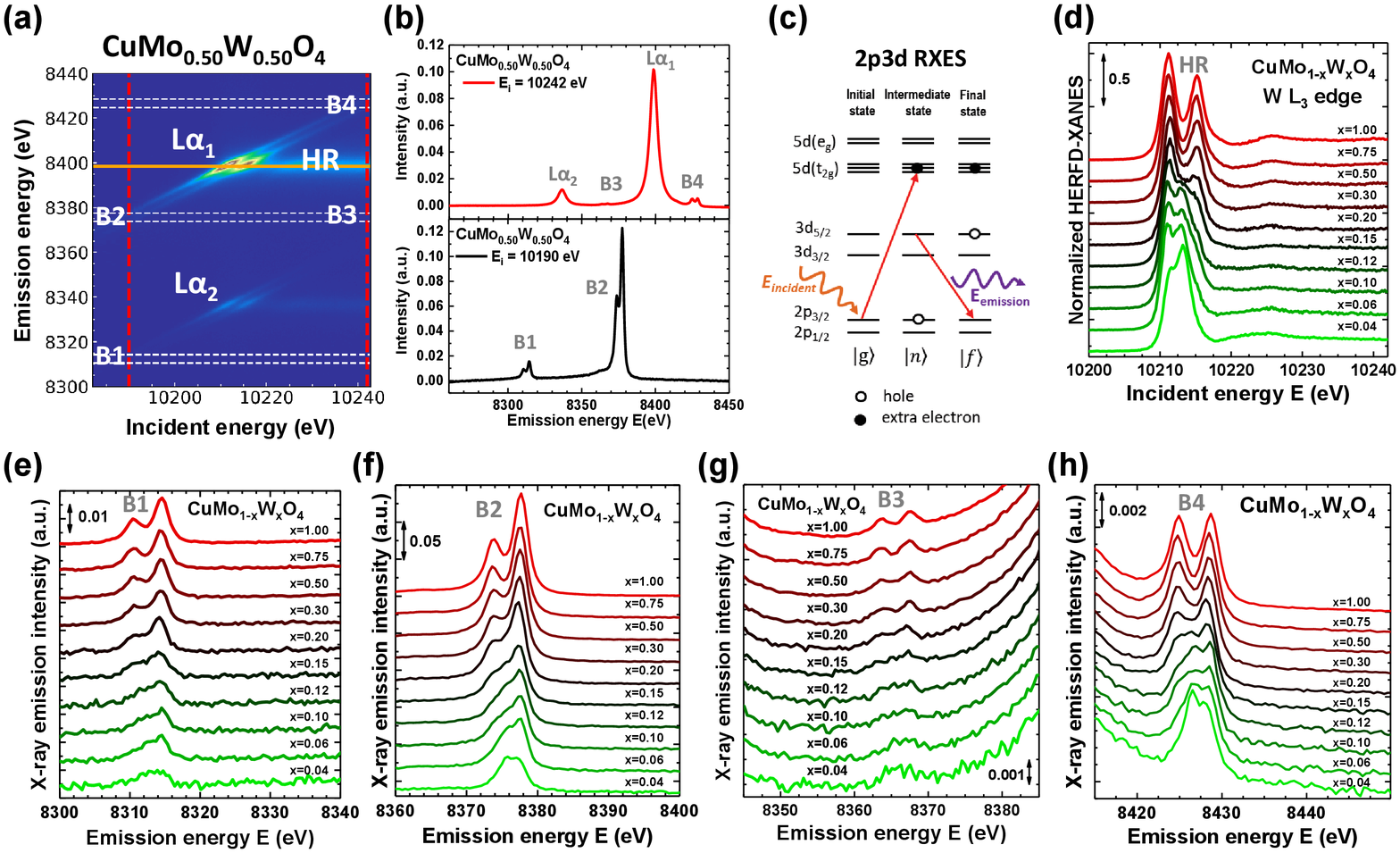}
	\caption{
		(a) RXES intensity map as a function of incident and emitted
		energies for CuMo$_{0.50}$W$_{0.50}$O$_4$ solid solution at 300~K. 
		(b) Vertical cuts of the RXES plane at the incident energies E$_i$=10190~eV and 10242~eV, shown in (a) by two vertical red dashed lines. Principal observed bands (L$\alpha_{1},\alpha_{2}$ and B1-B4) are labeled.  
		(c) A schematic diagram of 2p3d RXES process for octahedral tungsten coordination showing initial, intermediate and final states.
		(d) The W L$_3$-edge HERFD-XANES spectra for different CuMo$_{1-x}$W$_x$O$_4$ solid solutions measured at the emission energy E$_e$=8398.5$\pm$0.2~eV, indicated by the horizontal orange solid line in (a).		
		(e,f) High energy resolution off-resonant X-ray spectra for different  CuMo$_{1-x}$W$_x$O$_4$ solid solutions obtained at E$_i$=10190~eV below the W L$_3$-edge. 
		(g,h) High energy resolution off-resonant X-ray emission spectra for different  CuMo$_{1-x}$W$_x$O$_4$ solid solutions obtained at E$_i$=10242~eV above the W L$_3$-edge. }
	\label{fig3}
\end{figure*}

In the RXES experiment, the intensities and energies of the incoming and emitted X-rays are monitored. One way of presenting the RXES intensity plane is to plot the energy of emitted photons against the incident photon energy as shown in Fig.\ \ref{fig3}(a). Different plane cuts allow one to obtain valuable information on the structural and electronic properties of the material.  
For example, the W L$_3$-edge HERFD-XANES spectrum can be obtained from Fig.\ \ref{fig3}(a) by
integrating the emission intensity in the narrow emission energy region of $\pm$0.2~eV around 8398.5~eV. Thus extracted HERFD-XANES spectra of CuMo$_{1-x}$W$_x$O$_4$  solid solutions  at 300~K are shown in Fig.\ \ref{fig3}(d). 

Besides, the HERFD-XANES spectra, one can additionally obtain (Figs.\ \ref{fig3}(a,b)) high energy resolution off-resonant X-ray emission spectra by extracting them slightly below (bands B1 and B2 in Figs.\ \ref{fig3}(e,f)) and above (bands B3 and B4 in Figs.\ \ref{fig3}(g,h))  the W L$_3$ absorption edge, respectively. As one can see, 
both types of emission spectra contain information on the crystal field splitting, while with a smaller signal-to-noise ratio compared to HERFD-XANES (Fig.\ \ref{fig3}(d)) due to weaker intensity. Besides, in the non-resonant case (Figs.\ \ref{fig3}(g,h)), the tails of the main X-ray emission line (W L$_{\alpha_1}$ in Fig.\ \ref{fig3}(b)) produce some background under the B3 and B4 bands, which also limits the accuracy of the analysis and should be removed in advance.     

Further, we will discuss the effect of concentration and temperature on the local structure of tungsten ions in CuMo$_{1-x}$W$_x$O$_4$ solid solutions observed in the RXES spectra.

\subsection{Effect of composition}
The concentration dependences of the RXES spectra of CuMo$_{1-x}$W$_x$O$_4$ solid solutions at 300~K are shown in Figs.\ \ref{fig3}(d-h). 
The origin of the observed bands is due to three different mechanisms, already discussed above.

The emission spectra for the incident X-ray energy E$_i$=10190~eV  below the W L$_3$ absorption edge are shown in Figs.\ \ref{fig3}(e,f) and correspond to off-resonant condition  \cite{BLACHUCKI2017}. 
Two emission bands B1 at $\sim$8312~eV and B2 at $\sim$8376~eV can be distinguished in the spectra and correspond to the transitions from 3d$_{3/2}$ and 3d$_{5/2}$ levels, respectively (Figs.\ \ref{fig3}(b,c)). 
The intensity of the lower energy band B1 is smaller than that of the band B2, correlating with the intensity of the W L$\alpha_1$ and  L$\alpha_2$ bands. 

Both bands B1 and B2 are split into two peaks due to the crystal field of oxygen ligands. The distance between these peaks and the ratio $I_1 / I_2$ of their intensities were evaluated by fitting the spectra with two Lorentzian functions and were used to estimate the symmetry of the crystal field. The difference between the position of the two Lorentzian functions is related to the crystal-field splitting parameter $\Delta$ or 10Dq used in the crystal-field theory \cite{Bersuker2010}.

One can see in Figs.\ \ref{fig3}(e,f), that the change of the spectral shape upon tungsten content increase indicates that the local coordination of tungsten ions changes from tetrahedral ($x=0.04$) to octahedral ($x > 0.15$).

The emission spectra for the incident X-ray energy  E$_i$=10242~eV  well above the W L$_3$ absorption edge  are reported in Figs.\ \ref{fig3}(g,h).  In this case, four emission bands
W L$\alpha_2$, B3, W L$\alpha_1$ and B4 are observed in Fig.\ \ref{fig3}(b). 
The origin of the bands B3 and B4 is similar to that of B1 and B2, i.e., is due to the transitions from 3d$_{3/2}$ and 3d$_{5/2}$ levels, respectively (Fig.\ \ref{fig3}(c)).  
Again the lower energy band B3 has a much weaker intensity than that of the band B4, and both bands are split by the crystal field to 
two peaks (Figs.\ \ref{fig3}(g,h)). Opposite to the case of the bands B1 and B2, the bands B3 and B4 are located on top of the background due to the W L$\alpha_1$ and partially 
W L$\alpha_2$ fluorescence, which makes it more difficult to determine the magnitude of the peaks splitting and the ratio of their intensities. 
Nevertheless, the obtained parameters agree well with that obtained from the analysis of the bands  B1 and B2 suggesting the octahedral coordination of tungsten ions for $x > 0.15$. 

\begin{figure}[t]
	\centering
	\includegraphics[width=0.9\textwidth]{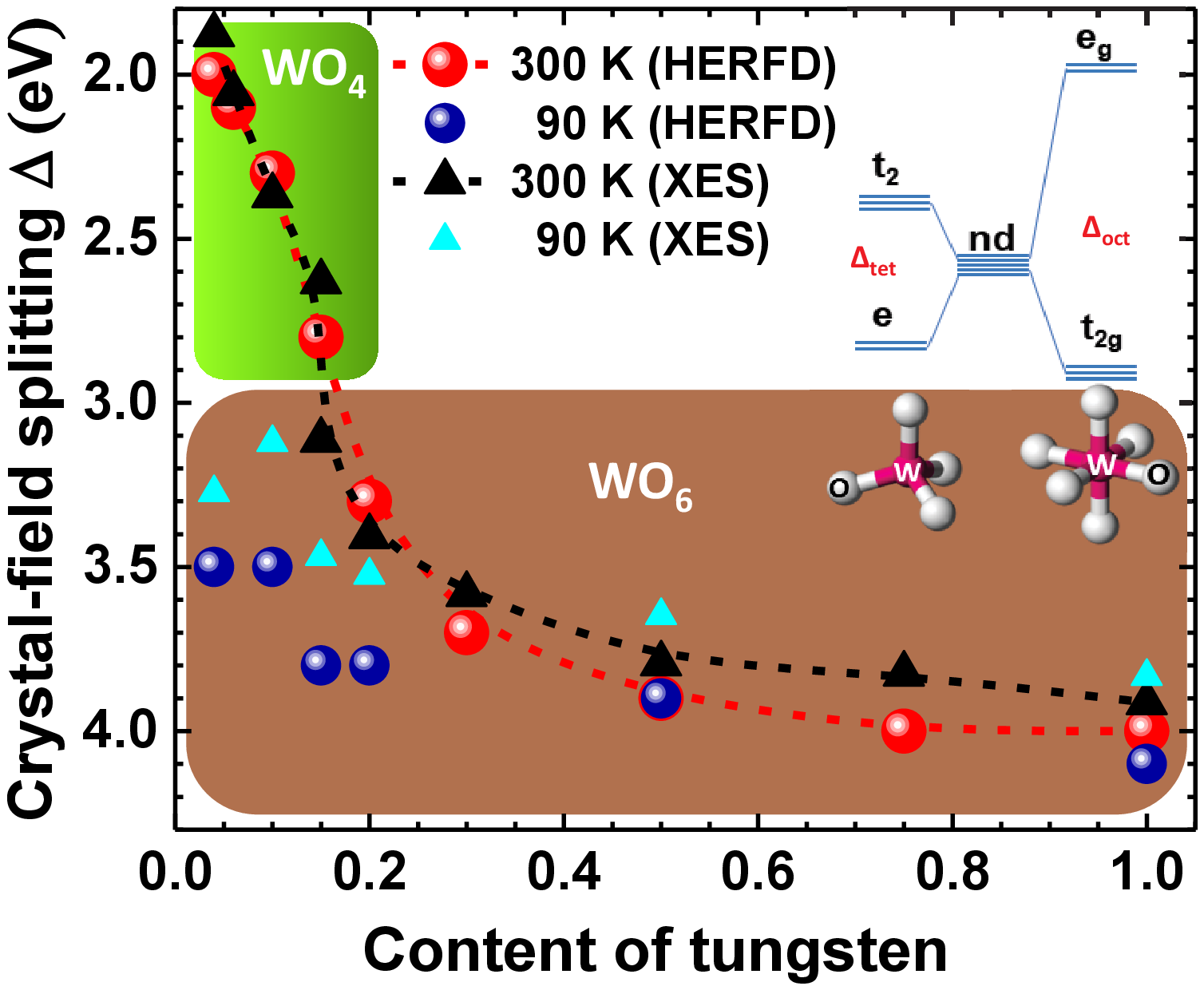}
	\caption{Crystal-field splitting parameter $\Delta$ as a function of tungsten content $x$ in CuMo$_{1-x}$W$_x$O$_4$ solid solutions at 90~K and 300~K calculated from off-resonant XES (Fig.\ \protect\ref{fig3}(f)) and HERFD-XANES (Fig.\ \protect\ref{fig3}(d)) spectra. Schematic energy level diagram showing crystal-field splitting of tungsten 5d-states in the tetrahedral and octahedral environment is also shown. Note that $\Delta_{tet}$ is smaller than $\Delta_{oct}$.}
	\label{fig4}
\end{figure}

Finally, the analysis was also performed for the W L$_3$-edge HERFD-XANES 
spectra (Fig.\ \ref{fig3}(d)). Here the pronounced splitting of 
the white line due to the octahedral crystal field is clearly visible for $x>0.15$
and agrees with the results obtained from X-ray fluorescence 
spectra in Figs.\ \ref{fig3}(e-h). 
The white line was fitted with two Lorentzian functions to estimate its splitting magnitude. 
The results for the crystal-field splitting parameter $\Delta$  obtained from the off-resonant XES and HERFD-XANES spectra of CuMo$_{1-x}$W$_x$O$_4$ solid solutions at 300~K and 90~K are combined in Fig.\ \ref{fig4}.
 
At 300~K, CuWO$_4$ ($x=1$) and CuMo$_{0.96}$W$_{0.04}$O$_4$ ($x=0.04$) have wolframite \cite{Forsyth1991} and $\alpha$-CuMoO$_4$ \cite{Wiesmann1997} phases, respectively, which correspond to the octahedral and tetrahedral coordination of tungsten ions.  Upon cooling down to 90~K, CuWO$_4$ remains always in the triclinic phase (space group $P\bar{1}$) \cite{WEITZEL1970}.  
At the same time, CuMo$_{0.96}$W$_{0.04}$O$_4$ transforms on cooling to 
$\gamma$-CuMoO$_4$ phase \cite{Gaudon2007a,Gaudon2007b} with distorted octahedral  coordination of Mo$^{6+}$ and W$^{6+}$ ions. 

For intermediate compositions of CuMo$_{1-x}$W$_x$O$_4$ solid solutions, a mixture of  $\alpha$-CuMoO$_4$ and $\gamma$-CuMoO$_4$ phases co-exists in the samples with the relative amounts depending on the tungsten content and temperature range  \cite{Gaudon2007a,Robertson2015,Benchikhi2017}. In particular, 
when samples are cooled from 300~K down to 90~K, a transition from tetrahedral-to-octahedral tungsten coordination is observed for $x<0.20$ but not for $x \geq 0.20$ (Fig.\ \ref{fig4}). 

Our theoretical XANES calculations for distorted WO$_6$ and WO$_4$ environments (Fig.\ \ref{fig2}) suggest a value of the parameter $\Delta \sim 3-4$~eV for octahedral coordination and about 2~eV for tetrahedral one. 
Therefore, taking into account the experimentally observed $\alpha$-to-$\gamma$ transition and theoretical predictions, two ranges of the crystal-field splitting parameter $\Delta$ can be identified (see green and brown regions in Fig.\ \ref{fig4}): the tetrahedral coordination of tungsten ions is expected for $\Delta < 3$, whereas the octahedral coordination for $\Delta > 3$.  Thus, the simple analysis of the experimental off-resonant XES and HERFD-XANES spectra can provide useful information on the local coordination of tungsten ions. In the next section, we will demonstrate the sensitivity of the parameter $\Delta$  to the hysteretic phase transitions in CuMo$_{1-x}$W$_x$O$_4$.    

Furthermore, we found that the parameter $\Delta$  correlates well with the solid solution  optical properties (their color). Namely, at 300~K, the samples with $x \leq 0.15$ have the crystal-field splitting parameter $\Delta < 3$~eV and greenish colour, whereas the samples with $x \geq 0.20$ have $\Delta > 3$~eV and brown colour. Note, that upon cooling of the samples with $x \leq 0.15$ down to 90~K, their color changes to brown, in agreement with previous studies \cite{Wiesmann1997}, and  the parameter  $\Delta$ increases above 3~eV   indicating the transition to the phase with the octahedral tungsten coordination.

\subsection{Effect of temperature}

It is known that the first-order structural phase transition between $\alpha$ and $\gamma$ phases in CuMo$_{1-x}$W$_x$O$_4$ solid solutions for $x < 0.15$ has strong 
hysteretic behaviour, which can be controlled by the tungsten content \cite{Gaudon2007a,Robertson2015}. 
The hysteresis loop was first experimentally detected by optical, 
calorimetry and magnetic measurement \cite{Gaudon2007a}. 

In our previous studies \cite{Jonane2020,Jonane2018a}, we have shown for 
CuMoO$_4$ and CuMo$_{0.90}$W$_{0.10}$O$_4$ that the Mo K-edge XANES is sensitive to the $\alpha$-to-$\gamma$ phase transition due to the significant change in the local coordination of molybdenum ions from tetrahedral to octahedral one. Moreover, the analysis of the Mo K-edge 
XANES allowed us to determine the  transition hysteresis loop and 
to estimate the fraction of $\alpha$-phase upon cooling and heating. 
At the same time, the sensitivity of the Mo K-edge XANES to the $\alpha$-to-$\gamma$  transition is based on a variation of the pre-edge peak amplitude which is hidden by the natural broadening of the core level (5.8~eV at the Mo K-edge \cite{KESKIRAHKONEN1974}).
Therefore, the high resolution of the off-resonant XES and HERFD-XANES spectra (Fig.\ \ref{fig3}) should be more beneficial for tracking the $\alpha$-to-$\gamma$  phase transition.

\begin{figure}[t]
	\centering
	\includegraphics[width=0.9\textwidth]{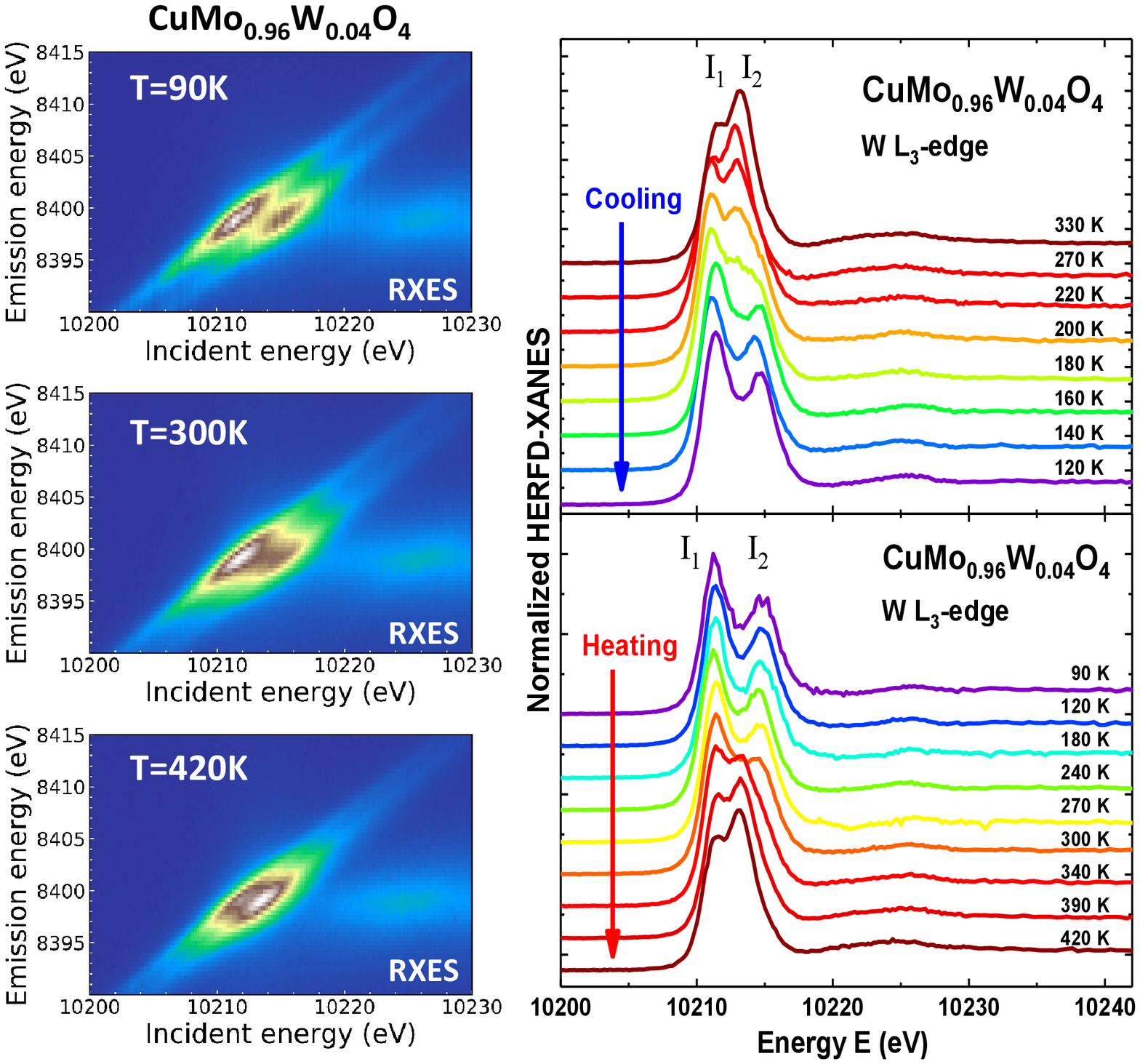}
	\caption{Parts of the RXES planes  at 90~K, 300~K and 420~K on heating (left panels) and temperature dependent W L$_3$-edge HERFD-XANES spectra measured on cooling and heating (right panel) for CuMo$_{0.96}$W$_{0.04}$O$_4$.}
	\label{fig5}
\end{figure}

\begin{figure}[t]
	\centering
	\includegraphics[width=0.6\textwidth]{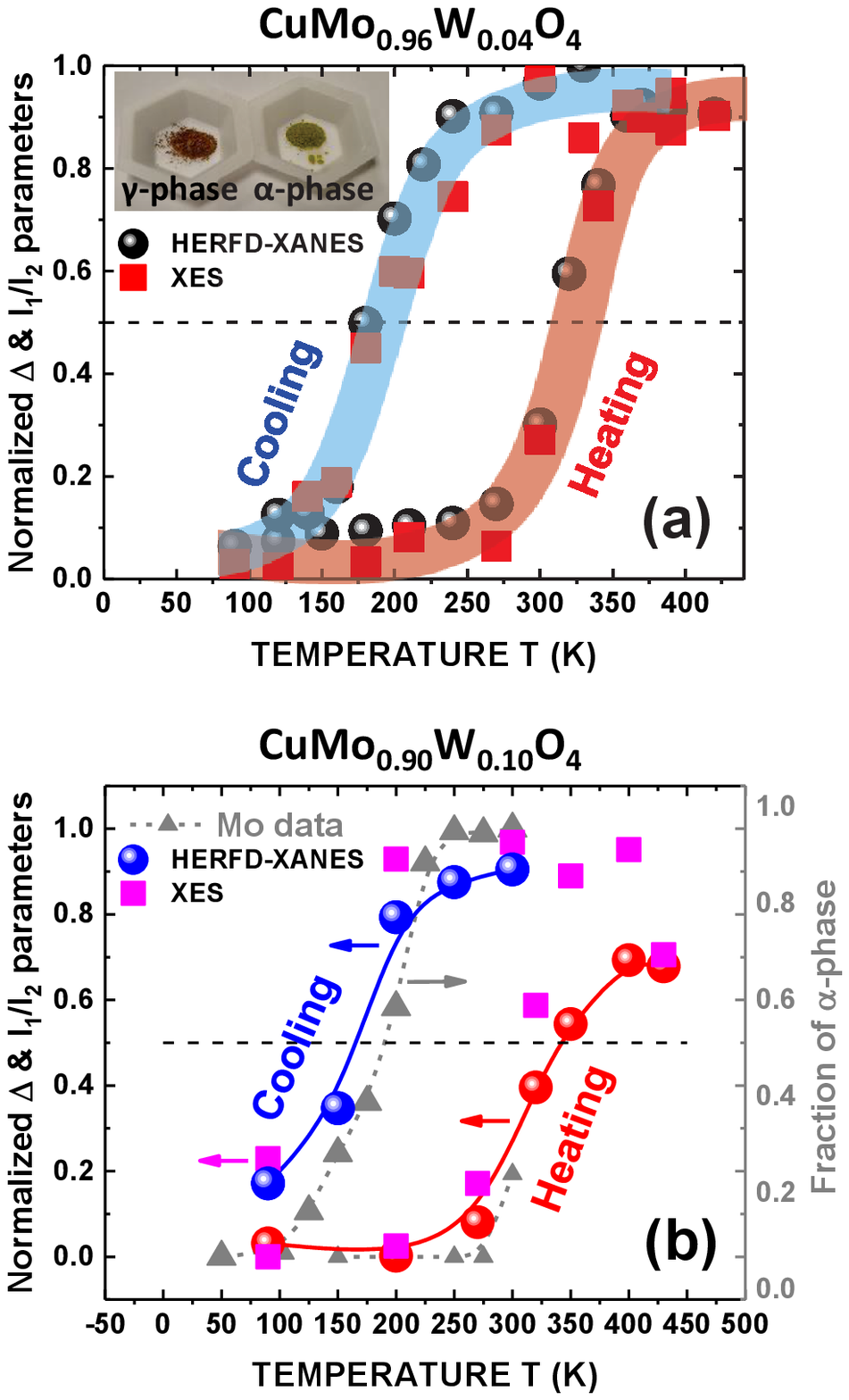}
	\caption{Temperature dependencies of the magnitude of the peak splitting ($\Delta$) and the ratio of the white line peak intensities ($I_1 / I_2$) for  CuMo$_{0.96}$W$_{0.04}$O$_4$ (a) and CuMo$_{0.90}$W$_{0.10}$O$_4$ (b). 
	A fraction of the $\alpha$-phase extracted from the analysis of the Mo K-edge XANES spectra of CuMo$_{0.90}$W$_{0.10}$O$_4$ \protect\cite{Jonane2020} is also displayed for comparison (solid triangles). Inset shows a photo of two CuMo$_{0.96}$W$_{0.04}$O$_4$ samples at room temperature: the brown sample is in the $\gamma$-phase (after treatment at 77~K in liquid nitrogen), and the green sample is in the $\alpha$-phase. }
	\label{fig6}
\end{figure}

An example of the temperature-dependent W L$_3$-edge HERFD-XANES spectra
of CuMo$_{0.96}$W$_{0.04}$O$_4$ is shown in Fig.\ \ref{fig5}. 
The change of the spectral shape reflects a transition from 
tetrahedral ($\alpha$-phase) to octahedral ($\gamma$-phase) tungsten  
coordination on cooling from 300~K to 120~K, whereas from octahedral to tetrahedral tungsten coordination on heating from 90~K to 420~K. 
Note that the effect is well observed in RXES planes as the temperature dependence of the relative position of two bright spots, which are 
well resolved in the octahedral coordination but change relative intensity and move closer in the tetrahedral one.

The phase transition hysteresis loops for CuMo$_{0.96}$W$_{0.04}$O$_4$ and CuMo$_{0.90}$W$_{0.10}$O$_4$ solid solutions are shown in Fig.\ \ref{fig6}.
They were determined from the temperature dependencies of the magnitude of the peak splitting and the ratio $I_1/ I_2$ of the peak intensities, which were 
obtained from the W L$_3$-edge HERFD-XANES and off-resonant XES spectra normalized on a scale from 0 to 1 to display them on the same graph.
Note that the two approaches have slightly different sensitivity to the change of tungsten coordination leading to some scatter of points.
The results for CuMo$_{0.90}$W$_{0.10}$O$_4$ (Fig.\ \ref{fig6}(b)) are also compared with a fraction of the $\alpha$-phase extracted from the analysis of the Mo K-edge XANES spectra reported in our previous work \cite{Jonane2020}. One can see a small 
difference between the starting temperatures of the $\alpha$-to-$\gamma$ phase transition obtained from the W L$_3$ and Mo K edges for CuMo$_{0.90}$W$_{0.10}$O$_4$ sample. When probed at the Mo K-edge, the transition starts at a slightly lower temperature than when measured at the W L$_3$ edge. This seeming contradiction has a simple explanation by the fact, that  some part of the tungsten atoms is located in the octahedral environment already at room temperature, as was concluded above from Fig.\ \ref{fig4}.

\section{Conclusions}
\label{conc}

Resonant X-ray emission spectroscopy at the W L$_3$-edge  was used to study the
changes in the local atomic structure of tungsten ions across the thermochromic phase transition in CuMo$_{1-x}$W$_x$O$_4$ solid solutions
as a function of sample composition and temperature.

We demonstrated that the analysis of the RXES plane provides useful information on the coordination of tungsten atoms in the sample bulk and allows one to determine the  crystal-field splitting parameter $\Delta$ for the 5d(W)-states. Moreover, this information can be extracted from the RXES plane using two different approaches by analysing the high-energy resolution fluorescence detected X-ray absorption near-edge structure (HERFD-XANES) and the high energy resolution off-resonant X-ray emission spectra excited below and above resonance conditions. 
The analysis of the RXES planes shows a clear advantage over conventional XANES due to revealing spectral features with much higher resolution.

The method is well suited for in-situ measurements and was used here to determine the hysteretic behaviour of  the first-order structural phase transition between $\alpha$ and $\gamma$ phases in CuMo$_{1-x}$W$_x$O$_4$ solid solutions on cooling and heating, even at low ($x < 0.10$) tungsten content.  

We found that tungsten ions in CuMo$_{1-x}$W$_x$O$_4$ solid solutions 
have octahedral coordination for $x > 0.15$ at all temperatures, whereas their coordination changes from tetrahedral to octahedral upon cooling for smaller tungsten content. Nevertheless, some amount of tungsten ions co-exists in the octahedral 
environment at room temperature for $x < 0.15$. 
The obtained results correlate well with the optical properties of these 
materials, in particular, color change from green to brown upon cooling 
or increasing tungsten content. 

The electronic structure of CuMo$_{1-x}$W$_x$O$_4$ solid solutions
controls their thermochromic properties, which are related to a variation of the transmission window in optical spectra \cite{Gaudon2007a,Wu2020}. A change in the local coordination of tungsten atoms from tetrahedral to octahedral affects the band gap, which is determined by the  oxygen-to-metal charge transfer \cite{Gaudon2007a}. The band gap is smaller in the case of octahedral coordination of tungsten \cite{Wu2020}, i.e., at  higher tungsten content or lower temperature.

To conclude, this study demonstrates the possibilities of the RXES technique to probe the crystal field effect in functional thermochromic materials with controllable properties on the example of CuMo$_{1-x}$W$_x$O$_4$ solid solutions.

\section*{Declaration of Competing Interest}

The authors declare that they have no known competing financial interests or personal relationships that could have appeared to influence the work reported in this paper.

\section*{Acknowledgements}

Financial support provided by Scientific Research Project for Students and Young Researchers Nr. SJZ/2019/1 realized at the Institute of Solid State Physics, University of Latvia is greatly acknowledged.
The used infrastructure of the von Hamos spectrometer was realized in the frame of projects FKZ 05K13UK1 and FKZ 05K14PP1.
The experiment at the PETRA III synchrotron was performed within the project No. I-20180615 EC. The synchrotron experiments have been supported by the project CALIPSOplus under the Grant Agreement 730872 from the EU Framework Programme for Research and Innovation HORIZON 2020.
The experiment at the Elettra synchrotron was performed within the project No. 20150303.
Institute of Solid State Physics, University of Latvia as the Center of Excellence has received funding from the European Union's Horizon 2020 Framework Programme H2020-WIDESPREAD-01-2016-2017-TeamingPhase2 under grant agreement No. 739508, project CAMART2.

\section*{Supplementary materials}

Supplementary material associated with this article can be found, in the online version, at doi: XXX.j.actamat.2020.YYY



\begin{thebibliography}{10}
	\expandafter\ifx\csname url\endcsname\relax
	\def\url#1{\texttt{#1}}\fi
	\expandafter\ifx\csname urlprefix\endcsname\relax\def\urlprefix{URL }\fi
	\expandafter\ifx\csname href\endcsname\relax
	\def\href#1#2{#2} \def\path#1{#1}\fi
	
	\bibitem{Gaudon2007a}
	M.~Gaudon, C.~Carbonera, A.~E. Thiry, A.~Demourgues, P.~Deniard, C.~Payen,
	J.~F. L\'etard, S.~Jobic, {Adaptable thermochromism in the
		CuMo$_{1-x}$W$_x$O$_4$ series ($0 \leq x < 0.1$): a behavior related to a
		first-order phase transition with a transition temperature depending on x},
	Inorg. Chem. 46 (2007) 10200--10207.
	\newblock \href {http://dx.doi.org/10.1021/ic701263c}
	{\path{doi:10.1021/ic701263c}}.
	
	\bibitem{Gaudon2007b}
	M.~Gaudon, P.~Deniard, A.~Demourgues, A.~E. Thiry, C.~Carbonera, A.~{Le
		Nestour}, A.~Largeteau, J.~F. L\'etard, S.~Jobic, {Unprecedented
		``one-finger-push''-induced phase transition with a drastic color change in
		an inorganic material}, Adv. Mater. 19 (2007) 3517--3519.
	\newblock \href {http://dx.doi.org/10.1002/adma.200700905}
	{\path{doi:10.1002/adma.200700905}}.
	
	\bibitem{Gaudon2009}
	M.~Gaudon, B.~Basly, Y.~Fauque, J.~Majimel, M.~H. Delville, {Thermochromic
		phase transition on CuMo$_{0.9}$W$_{0.1}$O$_4$@SiO$_2$ core-shell particles},
	Inorg. Chem. 48 (2009) 2136--2139.
	\newblock \href {http://dx.doi.org/10.1021/ic802057c}
	{\path{doi:10.1021/ic802057c}}.
	
	\bibitem{Yanase2013}
	I.~Yanase, T.~Mizuno, H.~Kobayashi, {Structural phase transition and
		thermochromic behavior of synthesized W-substituted CuMoO$_4$}, Ceram. Int.
	39 (2013) 2059--2064.
	\newblock \href {http://dx.doi.org/10.1016/j.ceramint.2012.08.059}
	{\path{doi:10.1016/j.ceramint.2012.08.059}}.
	
	\bibitem{Robertson2015}
	L.~Robertson, N.~Penin, V.~Blanco-Gutierrez, D.~Sheptyakov, A.~Demourgues,
	M.~Gaudon, {CuMo$_{0.9}$W$_{0.1}$O$_4$ phase transition with thermochromic,
		piezochromic, and thermosalient effects}, J. Mater. Chem. C 3 (2015)
	2918--2924.
	\newblock \href {http://dx.doi.org/10.1039/C4TC02463J}
	{\path{doi:10.1039/C4TC02463J}}.
	
	\bibitem{Wu2020}
	X.~Wu, C.~Fu, J.~Cao, C.~Gu, W.~Liu, {Effect of W doping on phase transition
		behavior and dielectric relaxation of CuMoO$_4$ obtained by a modified
		sol-gel method}, Mater. Res. Express 7 (2020) 016309.
	\newblock \href {http://dx.doi.org/10.1088/2053-1591/ab6546}
	{\path{doi:10.1088/2053-1591/ab6546}}.
	
	\bibitem{Blanco2015}
	V.~Blanco-Gutierrez, L.~Cornu, A.~Demourgues, M.~Gaudon,
	{CoMoO$_4$/CuMo$_{0.9}$W$_{0.1}$O$_4$ mixture as an efficient piezochromic
		sensor to detect temperature/pressure shock parameters}, ACS Appl. Mater.
	Interfaces 7 (2015) 7112--7117.
	\newblock \href {http://dx.doi.org/10.1021/am508652h}
	{\path{doi:10.1021/am508652h}}.
	
	\bibitem{Gaudon2010}
	M.~Gaudon, C.~Riml, A.~Turpain, C.~Labrugere, M.~H. Delville, {Investigation of
		the chromic phase transition of CuMo$_{0.9}$W$_{0.1}$O$_4$ induced by surface
		protonation}, Chem. Mater. 22 (2010) 5905--5911.
	\newblock \href {http://dx.doi.org/10.1021/cm101824d}
	{\path{doi:10.1021/cm101824d}}.
	
	\bibitem{Hill2013}
	J.~C. Hill, Y.~Ping, G.~A. Galli, K.-S. Choi, {Synthesis, photoelectrochemical
		properties, and first principles study of n-type CuW$_{1-x}$Mo$_x$O$_4$
		electrodes showing enhanced visible light absorption}, Energy Environ. Sci. 6
	(2013) 2440--2446.
	\newblock \href {http://dx.doi.org/10.1039/C3EE40827B}
	{\path{doi:10.1039/C3EE40827B}}.
	
	\bibitem{Liang2018}
	Q.~Liang, Y.~Guo, N.~Zhang, Q.~Qian, Y.~Hu, J.~Hu, Z.~Li, Z.~Zou, {Improved
		water-splitting performances of CuW$_{1-x}$Mo$_x$O$_4$ photoanodes
		synthesized by spray pyrolysis}, Sci. China Mater. 61 (2018) 1297--1304.
	\newblock \href {http://dx.doi.org/10.1007/s40843-018-9287-5}
	{\path{doi:10.1007/s40843-018-9287-5}}.
	
	\bibitem{Joseph2020}
	N.~Joseph, J.~Varghese, M.~Teirikangas, H.~Jantunen, A temperature-responsive
	copper molybdate polymorph mixture near to water boiling point by a simple
	cryogenic quenching route, ACS Appl. Mater. Interfaces 12 (2020) 1046--1053.
	\newblock \href {http://dx.doi.org/10.1021/acsami.9b17300}
	{\path{doi:10.1021/acsami.9b17300}}.
	
	\bibitem{Steiner2001}
	G.~Steiner, R.~Salzer, W.~Reichelt, {Temperature dependence of the optical
		properties of CuMoO$_4$}, Fresenius J. Anal. Chem. 370 (2001) 731--734.
	\newblock \href {http://dx.doi.org/10.1007/s002160000630}
	{\path{doi:10.1007/s002160000630}}.
	
	\bibitem{Jonane2019}
	I.~Jonane, A.~Anspoks, G.~Aquilanti, A.~Kuzmin, {High-temperature X-ray
		absorption spectroscopy study of thermochromic copper molybdate}, Acta Mater.
	179 (2019) 26--35.
	\newblock \href {http://dx.doi.org/10.1016/j.actamat.2019.06.034}
	{\path{doi:10.1016/j.actamat.2019.06.034}}.
	
	\bibitem{Bordiga2013}
	S.~Bordiga, E.~Groppo, G.~Agostini, J.~A. van Bokhoven, C.~Lamberti,
	{Reactivity of surface species in heterogeneous catalysts probed by In situ
		X-ray absorption techniques}, Chem. Rev. 113 (2013) 1736--1850.
	\newblock \href {http://dx.doi.org/10.1021/cr2000898}
	{\path{doi:10.1021/cr2000898}}.
	
	\bibitem{YOUNG2014}
	N.~A. Young, {The application of synchrotron radiation and in particular X-ray
		absorption spectroscopy to matrix isolated species}, Coord. Chem. Rev.
	277-278 (2014) 224--274.
	\newblock \href {http://dx.doi.org/10.1016/j.ccr.2014.05.010}
	{\path{doi:10.1016/j.ccr.2014.05.010}}.
	
	\bibitem{Mastelaro2018}
	V.~R. Mastelaro, E.~D. Zanotto, {X-ray absorption fine structure (XAFS) studies
		of oxide glasses-A 45-year overview}, Materials 11 (2018) 204.
	\newblock \href {http://dx.doi.org/10.3390/ma11020204}
	{\path{doi:10.3390/ma11020204}}.
	
	\bibitem{Timoshenko2015}
	J.~Timoshenko, A.~Anspoks, A.~Kalinko, I.~Jonane, A.~Kuzmin, {Local structure
		of multiferroic MnWO$_4$ and Mn$_{0.7}$Co$_{0.3}$WO$_4$ revealed by the
		evolutionary algorithm}, Ferroelectrics 483 (2015) 68--74.
	\newblock \href {http://dx.doi.org/10.1080/00150193.2015.1058687}
	{\path{doi:10.1080/00150193.2015.1058687}}.
	
	\bibitem{Kalinko2016}
	A.~Kalinko, M.~Bauer, J.~Timoshenko, A.~Kuzmin, {Molecular dynamics and reverse
		Monte Carlo modeling of scheelite-type AWO$_4$ (A= Ca, Sr, Ba) W L$_3$-edge
		EXAFS spectra}, Phys. Scr. 91 (2016) 114001.
	\newblock \href {http://dx.doi.org/10.1088/0031-8949/91/11/114001}
	{\path{doi:10.1088/0031-8949/91/11/114001}}.
	
	\bibitem{Timoshenko2016}
	J.~Timoshenko, A.~Anspoks, A.~Kalinko, A.~Kuzmin, {Local structure of cobalt
		tungstate revealed by EXAFS spectroscopy and reverse Monte Carlo/evolutionary
		algorithm simulations}, Z. Phys. Chem. 230 (2016) 551--568.
	\newblock \href {http://dx.doi.org/10.1515/zpch-2015-0646}
	{\path{doi:10.1515/zpch-2015-0646}}.
	
	\bibitem{Kuzmin2020}
	A.~Kuzmin, J.~Timoshenko, A.~Kalinko, I.~Jonane, A.~Anspoks, {Treatment of
		disorder effects in X-ray absorption spectra beyond the conventional
		approach}, Rad. Phys. Chem. 175 (2020) 108112.
	\newblock \href {http://dx.doi.org/10.1016/j.radphyschem.2018.12.032}
	{\path{doi:10.1016/j.radphyschem.2018.12.032}}.
	
	\bibitem{Jonane2020}
	I.~Jonane, A.~Cintins, A.~Kalinko, R.~Chernikov, A.~Kuzmin, {Low temperature
		X-ray absorption spectroscopy study of CuMoO$_4$ and
		CuMo$_{0.90}$W$_{0.10}$O$_4$ using reverse Monte-Carlo method}, Rad. Phys.
	Chem. 175 (2020) 108411.
	
	\bibitem{Yamazoe2008}
	S.~Yamazoe, Y.~Hitomi, T.~Shishido, T.~Tanaka, {XAFS study of tungsten L$_1$-
		and L$_3$-edges: structural analysis of WO$_3$ species loaded on TiO$_2$ as a
		catalyst for photo-oxidation of NH$_3$}, J. Phys. Chem. C 112 (2008)
	6869--6879.
	\newblock \href {http://dx.doi.org/10.1021/jp711250f}
	{\path{doi:10.1021/jp711250f}}.
	
	\bibitem{Jayarathne2014}
	U.~Jayarathne, P.~Chandrasekaran, A.~F. Greene, J.~T. Mague, S.~DeBeer, K.~M.
	Lancaster, S.~Sproules, J.~P. Donahue, {X-ray absorption spectroscopy
		systematics at the tungsten L-edge}, Inorg. Chem. 53 (2014) 8230--8241.
	\newblock \href {http://dx.doi.org/10.1021/ic500256a}
	{\path{doi:10.1021/ic500256a}}.
	
	\bibitem{KESKIRAHKONEN1974}
	O.~Keski-Rahkonen, M.~O. Krause, {Total and partial atomic-level widths}, At.
	Data Nucl. Data Tables 14 (1974) 139--146.
	\newblock \href {http://dx.doi.org/10.1016/S0092-640X(74)80020-3}
	{\path{doi:10.1016/S0092-640X(74)80020-3}}.
	
	\bibitem{BALERNA1991}
	A.~Balerna, E.~Bernieri, E.~Burattini, A.~Kuzmin, A.~Lusis, J.~Purans,
	P.~Cikmach, {XANES studies of MeO$_{3-x}$ (Me = W, Re, Ir) crystalline and
		amorphous oxides}, Nucl. Instrum. Methods Phys. Res. A 308 (1991) 240--242.
	\newblock \href {http://dx.doi.org/10.1016/0168-9002(91)90637-6}
	{\path{doi:10.1016/0168-9002(91)90637-6}}.
	
	\bibitem{Kuzmin1997spie}
	A.~Kuzmin, J.~Purans, {X-ray absorption spectroscopy study of the local
		environment around tungsten and molybdenum ions in tungsten-phosphate and
		molybdenum-phosphate glasses}, Proc. SPIE 2968 (1997) 180--185.
	\newblock \href {http://dx.doi.org/10.1117/12.266831}
	{\path{doi:10.1117/12.266831}}.
	
	\bibitem{Charton2002}
	P.~Charton, L.~Gengembre, P.~Armand, {TeO$_2$--WO$_3$ glasses: infrared, XPS
		and XANES structural characterizations}, J. Solid State Chem. 168 (2002)
	175--183.
	\newblock \href {http://dx.doi.org/10.1006/jssc.2002.9707}
	{\path{doi:10.1006/jssc.2002.9707}}.
	
	\bibitem{PETRAP64}
	W.~A. Caliebe, V.~Murzin, A.~Kalinko, M.~G{\"o}rlitz, {High-flux XAFS-beamline
		P64 at PETRA III}, AIP Conf. Proc. 2054 (2019) 060031.
	\newblock \href {http://dx.doi.org/10.1063/1.5084662}
	{\path{doi:10.1063/1.5084662}}.
	
	\bibitem{HAMOS2020}
	A.~Kalinko, W.~A. Caliebe, R.~Schoch, M.~Bauer, {A von Hamos-type hard X-ray
		spectrometer at the PETRA III beamline P64}, J. Synchrotron Rad. 27 (2020)
	31--36.
	\newblock \href {http://dx.doi.org/10.1107/S1600577519013638}
	{\path{doi:10.1107/S1600577519013638}}.
	
	\bibitem{Di2009}
	A.~{Di Cicco}, G.~Aquilanti, M.~Minicucci, E.~Principi, N.~Novello,
	A.~Cognigni, L.~Olivi, {Novel XAFS capabilities at ELETTRA synchrotron light
		source}, J. Phys.: Conf. Ser. 190 (2009) 012043.
	\newblock \href {http://dx.doi.org/10.1088/1742-6596/190/1/012043}
	{\path{doi:10.1088/1742-6596/190/1/012043}}.
	
	\bibitem{Joly2001}
	Y.~Joly, {X-ray absorption near-edge structure calculations beyond the
		muffin-tin approximation}, Phys. Rev. B 63 (2001) 125120.
	\newblock \href {http://dx.doi.org/10.1103/PhysRevB.63.125120}
	{\path{doi:10.1103/PhysRevB.63.125120}}.
	
	\bibitem{Bunuau2009}
	O.~Bun{\u{a}}u, Y.~Joly, Self-consistent aspects of x-ray absorption
	calculations, J. Phys.: Condens. Matter 21 (2009) 345501.
	\newblock \href {http://dx.doi.org/10.1088/0953-8984/21/34/345501}
	{\path{doi:10.1088/0953-8984/21/34/345501}}.
	
	\bibitem{Hedin1971}
	L.~Hedin, B.~I. Lundqvist, {Explicit local exchange-correlation potentials}, J.
	Phys. C: Solid State Phys. 4 (1971) 2064--2083.
	\newblock \href {http://dx.doi.org/10.1088/0022-3719/4/14/022}
	{\path{doi:10.1088/0022-3719/4/14/022}}.
	
	\bibitem{Wiesmann1997}
	M.~Wiesmann, H.~Ehrenberg, G.~Miehe, T.~Peun, H.~Weitzel, H.~Fuess, {$p$-$T$
		phase diagram of CuMoO$_4$}, J. Solid State Chem. 132 (1997) 88--97.
	\newblock \href {http://dx.doi.org/10.1006/jssc.1997.7413}
	{\path{doi:10.1006/jssc.1997.7413}}.
	
	\bibitem{Forsyth1991}
	J.~Forsyth, C.~Wilkinson, A.~Zvyagin, {The antiferromagnetic structure of
		copper tungstate, CuWO$_4$}, J. Phys.: Condens. Matter 3 (1991) 8433.
	\newblock \href {http://dx.doi.org/10.1088/0953-8984/3/43/010}
	{\path{doi:10.1088/0953-8984/3/43/010}}.
	
	\bibitem{Shannon1976}
	R.~D. Shannon, Revised effective ionic radii and systematic studies of
	interatomic distances in halides and chalcogenides, Acta Cryst. A 32 (1976)
	751--767.
	\newblock \href {http://dx.doi.org/10.1107/S0567739476001551}
	{\path{doi:10.1107/S0567739476001551}}.
	
	\bibitem{Bersuker2010}
	I.~B. Bersuker, Electronic Structure and Properties of Transition Metal
	Compounds: Introduction to the Theory, 2nd Edition, John Wiley \& Sons,
	Hoboken, New Jersey, 2010.
	\newblock \href {http://dx.doi.org/10.1002/9780470573051}
	{\path{doi:10.1002/9780470573051}}.
	
	\bibitem{Brown1977}
	M.~Brown, R.~E. Peierls, E.~A. Stern, {White lines in x-ray absorption}, Phys.
	Rev. B 15 (1977) 738--744.
	\newblock \href {http://dx.doi.org/10.1103/PhysRevB.15.738}
	{\path{doi:10.1103/PhysRevB.15.738}}.
	
	\bibitem{Kramers1925}
	H.~A. Kramers, W.~Heisenberg, \"{U}ber die streuung von strahlung durch atome,
	Z. Physik 31 (1925) 681--708.
	\newblock \href {http://dx.doi.org/10.1007/BF02980624}
	{\path{doi:10.1007/BF02980624}}.
	
	\bibitem{Glatzel2005}
	P.~Glatzel, U.~Bergmann, {High resolution 1s core hole X-ray spectroscopy in 3d
		transition metal complexes -- electronic and structural information}, Coord.
	Chem. Rev. 249 (2005) 65--95.
	\newblock \href {http://dx.doi.org/10.1016/j.ccr.2004.04.011}
	{\path{doi:10.1016/j.ccr.2004.04.011}}.
	
	\bibitem{Rueff2010}
	J.-P. Rueff, A.~Shukla, {Inelastic x-ray scattering by electronic excitations
		under high pressure}, Rev. Mod. Phys. 82 (2010) 847--896.
	\newblock \href {http://dx.doi.org/10.1103/RevModPhys.82.847}
	{\path{doi:10.1103/RevModPhys.82.847}}.
	
	\bibitem{Glatzel2009}
	P.~Glatzel, M.~Sikora, G.~Smolentsev, M.~Fern\'{a}ndez-Garc\'{i}a, {Hard X-ray
		photon-in photon-out spectroscopy}, Catal. Today 145 (2009) 294--299.
	\newblock \href {http://dx.doi.org/10.1016/j.cattod.2008.10.049}
	{\path{doi:10.1016/j.cattod.2008.10.049}}.
	
	\bibitem{RUEFF2013}
	J.-P. Rueff, A.~Shukla, {A RIXS cookbook: Five recipes for successful RIXS
		applications}, J. Electron Spectrosc. Relat. Phenom. 188 (2013) 10--16.
	\newblock \href {http://dx.doi.org/10.1016/j.elspec.2013.04.014}
	{\path{doi:10.1016/j.elspec.2013.04.014}}.
	
	\bibitem{Rovezzi2014}
	M.~Rovezzi, P.~Glatzel, Hard x-ray emission spectroscopy: a powerful tool for
	the characterization of magnetic semiconductors, Semicond. Sci. Technol.
	29~(2) (2014) 023002.
	\newblock \href {http://dx.doi.org/10.1088/0268-1242/29/2/023002}
	{\path{doi:10.1088/0268-1242/29/2/023002}}.
	
	\bibitem{Wach2020}
	A.~Wach, J.~S{\'{a}}, J.~Szlachetko, {Comparative study of the around-Fermi
		electronic structure of 5$d$ metals and metal-oxides by means of
		high-resolution X-ray emission and absorption spectroscopies}, J. Synchrotron
	Rad. 27 (2020) 689--694.
	\newblock \href {http://dx.doi.org/10.1107/S1600577520003690}
	{\path{doi:10.1107/S1600577520003690}}.
	
	\bibitem{BLACHUCKI2017}
	W.~B\l{}achucki, J.~Hoszowska, J.-C. Dousse, Y.~Kayser, R.~Stachura,
	K.~Tyra\l{}a, K.~Wojtaszek, J.~S\'{a}, J.~Szlachetko, {High energy resolution
		off-resonant spectroscopy: A review}, Spectrochimica Acta B 136 (2017)
	23--33.
	\newblock \href {http://dx.doi.org/10.1016/j.sab.2017.08.002}
	{\path{doi:10.1016/j.sab.2017.08.002}}.
	
	\bibitem{WEITZEL1970}
	von {Hans Weitzel}, {Magnetische struktur von CoWO$_4$, NiWO$_4$ und CuWO$_4$},
	Solid State Commun. 8 (1970) 2071--2072.
	\newblock \href {http://dx.doi.org/10.1016/0038-1098(70)90221-8}
	{\path{doi:10.1016/0038-1098(70)90221-8}}.
	
	\bibitem{Benchikhi2017}
	M.~Benchikhi, R.~El~Ouatib, S.~Guillemet-Fritsch, L.~Er-Rakho, B.~Durand,
	{Investigation of structural transition in molybdates CuMo$_{1-x}$W$_x$O$_4$
		prepared by polymeric precursor method}, Process. Appl. Ceram. 11 (2017)
	21--26.
	\newblock \href {http://dx.doi.org/10.2298/PAC1701021B}
	{\path{doi:10.2298/PAC1701021B}}.
	
	\bibitem{Jonane2018a}
	I.~Jonane, A.~Cintins, A.~Kalinko, R.~Chernikov, A.~Kuzmin, {X-ray absorption
		near edge spectroscopy of thermochromic phase transition in CuMoO$_4$}, Low
	Temp. Phys. 44 (2018) 434--437.
	\newblock \href {http://dx.doi.org/10.1063/1.5034155}
	{\path{doi:10.1063/1.5034155}}.
	
\end{thebibliography}

\end{document}